# Cumulative, Adaptive, Open-ended Change through Self-Other Reorganization

## Reply to comment on 'An evolutionary process without variation and selection'

LIANE GABORA AND MIKE STEEL

**Abstract.** Self-Other Reorganization (SOR) is a theory of how interacting entities or individuals, each of which can be described as an autocatalytic network, collectively exhibit cumulative, adaptive, open-ended change, or evolution. Zachar et al.'s [61] critique of SOR stems from misunderstandings; it does not weaken the arguments in [27]. The formal framework of Reflexively Autocatalytic and foodset-derived sets (RAFs) enables us to model the process whereby, through their interactions, a set of elements become a 'collective self'. SOR shows how the RAF setting provides a means of encompassing abiogenesis and cultural evolution under the same explanatory framework and provides a plausible explanation for the origins of both evo- lutionary processes. Although SOR allows for detrimental stimuli (and products), there is (naturally) limited opportunity for elements that do not contribute to or reinforce a RAF to become part of it. Replication and cumulative, adaptive change in RAFs is well-established in the literature. Contrary to Zachar et al., SOR is not a pure percolation model (such as SIR); it encompasses not only learning (modeled as assimilation of foodset elements) but also creative restructuring (modeled as generation of foodset-derived elements), as well as the emergence of new structures made possible by new foodset- and foodset-derived elements. Cultural SOR is robust to degradation, and imperfect replication. Zachar et al.'s simulation contains no RAFs, and does not model SOR.

Liane Gabora (Corresponding Author)
Department of Psychology, University of British Columbia, Kelowna BC, Canada
liane.gabora@ubc.ca

Mike Steel
Biomathematics Research Centre, University of Canterbury, Christchurch, New Zealand

# 1. Introduction

Our recent paper [27] analysed how cumulative adaptive change in Reflexively Autocatalytic and Foodset-derived set (RAF) networks compares to natural selection, gave RAF evolution a name—Self-Other Reorganization (SOR)—and showed how SOR enables us to encompass abiogenesis and cultural evolution under the same explanatory framework. Zachar et al. [61] disagree with our use of the word 'evolution,' and interpret our framework as implying that SOR (a) assumes all changes are adaptive, and (b) does not allow for death and degradation. This reply addresses these claims, as well as other issues raised in Zachar et al.'s [61] commentary, including why SOR is not a pure percolation model, the relationship of SOR to learning, growth, and imperfect replication, and why SOR is modeled as a population-level (as opposed to individual-level) process.

# 2. Establishing a Common Terminology and Basic Framework

We begin by clearing up misunderstandings that are a matter of terminology, as opposed to conceptual error. Although terms were defined and used consistently in [27], given the interdisciplinary nature of the topic, extra care must be taken to ensure that everyone is on the same page.

2.1. **Evolution.** Zachar et al.'s defence of the theory of natural selection[1] suggests that they read our paper as disputing or challenging the theory of natural selection. This is not the case, as is made clear in the opening paragraph of the original manuscript: "Natural selection has provided the foundation of modern biology, and a unifying principle for describing species change" (p. 1).

---

[1]For example, they write, "There is an overwhelming consensus that the theory of evolution by natural selection of replicating and competing entities capable of multiplication heredity and variability is the adequate explanation for unsupervised adaptive change supported by innumerable studies and experiments on long-term evolution and adaptation of various species."

Zachar et al. accuse us of "using 'evolution' without disclosing right away the authors' meaning." We respectfully disagree; we state precisely what we mean by 'evolution' in the very first sentence of the paper: "Evolution refers to cumulative, adaptive change over time; thus, an evolutionary process produces increasingly adapted entities." Nor did we 'redefine' evolution, the word existed before Darwin used it to refer to natural selection, and epigenetic change is commonly conceived of as evolutionary.[2]

Since Zachar et al. make clear they prefer to reserve use of the word 'evolution' to refer to natural selection, for the sake of establishing a common terminology, we will avoid the term evolution as much as possible in this reply, and use instead the phrase, *cumulative, adaptive change*.

2.2. **Acquired Change versus Genomic Change.** When Zachar et al. speak of "inheritance of acquired mutations (changes) at the level of the genome" it becomes clear that for them the term 'acquired' includes genomic change due to mutation, whereas in our paper the term 'acquired' did not include genomic change due to mutation. Much misunderstanding arises from this simple difference in terminology. Zachar et al. claim that we falsely imply that "adaptations are never results of acquired traits," but we agree that if the term 'acquired change' is taken to encompass genomic change due to mutation, then adaptations can be the result of acquired traits.

Zachar et al. also accuse us of falsely implying that "acquired traits only accumulate when there is no germline/soma distinction." Once again, if the term 'acquired change' is taken to encompass genomic change due to mutation, then we agree that acquired traits can accumulate when there is a germline/soma distinction. So again, we are conceptually on the same page here.

---

[2]We note also that the term 'evolution' is used by cultural evolution researchers to refer to cumulative adaptive, open-ended change, and by physicists to refer to change in the quantum state of a system over time. In the current era of interdisciplinary science, progress cannot be made if we each assume our preferred definition is correct and all others are false.

While we recognize that one could consider genomic change due to mutation as 'acquired,' we find it more parsimonious to reserve the term *acquired change* for non-genomic change obtained during the lifetime and discarded at the end of the lifetime. This enables us to distinguish non-genomic acquired change from genomic change due to mutation, which we refer to simply as *genomic change*.

Having established how we use the term 'acquired change', and our rationale for this terminology, we restate our point: to the extent that acquired change (*not* genomic change) is heritable (vertically transmitted), the process does not obey the assumptions that make natural selection a valid model. The greater the degree to which acquired change is vertically transmitted, the more quickly acquired change drowns out the multi-generational process of natural selection.

2.3. **Reflexively Autocatalytic and Foodset-derived set (RAF).** There is considerable misunderstanding throughout the commentary about what a RAF is. As far as we can see, nothing in the commentary—including the figures and model—actually pertains to or describes or depicts a RAF.

To explain what a RAF is, we must first introduce the concept of a *catalytic reaction system* (CRS). A CRS is a tuple $Q = (X, R, C, F)$ consisting of a set $X$ of element types, a set $R$ of reactions, a catalysis set $C$ indicating which element types catalyse which reactions, and a subset $F$ of $X$ called the *foodset*. A *Reflexively Autocatalytic and F-derived* set—i.e., a RAF—is a non-empty subset $R' \subseteq R$ of reactions that satisfies the following properties:

(1) *Reflexively autocatalytic (RA)*: each reaction $r \in R'$ is catalysed by at least one element type that is either produced by $R'$ or present in the foodset $F$; and

(2) *F-derived*: all reactants in $R'$ can be generated from the foodset $F$ by using a series of reactions from $R'$ itself.

A set of reactions that forms a RAF is collectively autocatalytic (by the RA condition) , since each of its reactions is catalysed somewhere in the RAF, and self-sustaining (by the F-derived condition). Note that in RAF theory, a catalyst needn't necessarily cause more of something to be produced (or at a faster rate); the catalyst is simply the impetus that prompts a reaction to proceed.

A CRS may contain no RAFs, or a vast network of them. If it contains multiple RAFs there is a unique maximal one that contains all the others, referred to as the *maxRAF*. The maxRAF may consist of subRAFs, which are often clustered and hierarchically structured. It may also contain co-RAFs: structures that are not RAFs on their own but form RAFs, when combined with certain other RAFs. RAFs can form spontaneously, and expand through the merger of subRAFs and co-RAFs. Dynamic RAF networks can include transient RAFs, which include elements (sometimes referred to as nodes) and/or reaction paths that are unstable. The number of RAFs in a CRS changes over time as reactions generate new products that in turn enable new reactions and reaction pathways, as illustrated in Figure 1. The generalized RAF setting is conducive to the development of efficient (i.e., polynomial time) algorithms for questions that are computationally intractable (i.e., NP-hard [48]). This means it is possible to analyze vastly complex networks in terms of their RAF structure, as, for example, in [59, 60].

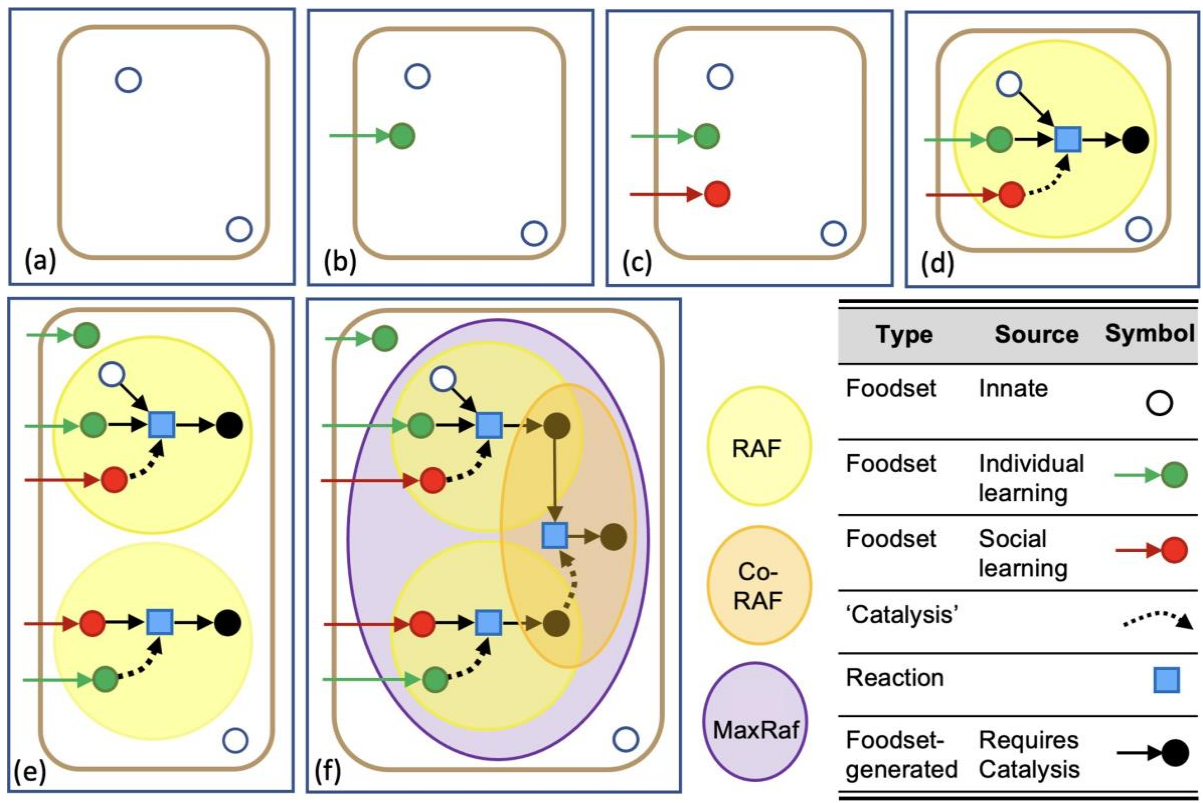

Figure 1. Example of cumulative network growth in a cognitive application of RAFs. Adapted from [20].

The original impetus behind the development of the RAF framework was to model the origin of life (OOL) [35, 36, 39, 38, 33, 34, 40, 50, 48, 49, 59], and the terminology used here reflects that initial area of application. That is, RAFs were first used to develop the hypothesis that life began as, not a single self-replicating molecule, but as a set of molecules that, through catalyzed reactions, collectively reconstituted the whole [42]. RAFs have also been used to model the emergence of cognitive structure capable of combining ideas, or adapting them to new situations, and thereby generating cumulative cultural change [20, 21, 24, 26, 25, 28, 31, 32]. When applied to cognition, the elements are mental representations, and the reactions are operations on mental representations (such as categorization or concept combination). External stimuli, internal drive states, and mental representations themselves play the role of catalysts, triggering conceptual change. In this paper, whether it be a biological CSR or a cultural CSR (i.e., a mind), for simplicity we may refer to it using the more familiar term, *individual*.

**2.4. Self-Other Reorgzanization (SOR).** Self-Other Reorgzanization (SOR) is a theory of how interacting entities or individuals, each of which constitutes a RAF network, collectively exhibit cumulative, adaptive change (what we call evolution). SOR differs from natural selec- tion in some key respects. First, SOR addresses the problem of origins, which (as noted by Darwin himself), natural selection does not address. Second, because there is no self-assembly code, SOR is a lower-fidelity process of adaptive change than natural selection. In SOR, cumu- lative, adaptive change arises due to the combined effect of RAF dynamics within individuals and communal exchange between individuals. SOR encompasses epigenetic change, but while epigenetic change refers to change that occurs in addition to genetic change to produce a pheno- type, cumulative adaptive change can take place through SOR in the absence of genes (or any other type of self-assembly code), and in the absence of any genotype-phenotype distinction.

### 3. Mis-characterization of SOR

Having established a common terminology and basic framework, we now turn to conceptual misunderstandings.

**3.1. Growth.** Zachar et al.'s claim that we "ignore growth" is incorrect. Growth was not the focus of this particular paper because the capacity for growth in RAF models had already been well established (see papers cited in Section 2.3). Because RAF elements modify network struc- ture, the RAF framework is consistent with the goal of understanding not just how networks are structured but also how they dynamically restructure themselves in response to internal and external pressures. Indeed, what differentiates RAFs from other models in network science is that the elements are not just passive transmitters of activation; they actively galvanize, or 'catalyze' the synthesis of novel (foodset-derived) elements from existing ones (the foodset).

This makes RAFs particularly suited to model how new structure grows out of earlier struc- ture, i.e., generative network growth [48], as seen in Figure 1. Such generativity may result in phase transitions to a network that is self-sustaining and self-organizing [35, 39, 43], as well as potentially able to evolve, i.e., exhibit cumulative, adaptive, open-ended change [38].

3.2. **Why Change in SOR is Adaptive.** Zachar et al. write, "without selection, there is no guarantee that maladaptive changes do not disrupt population members (RAFs), ultimately leading to the collapse of the population." We point out that there is also no guarantee against maladaptive changes wiping out a population that is evolving through natural selection. Zachar et al. write, "As there is no guarantee that these changes are adaptive and there is no mechanism to prevent non-adaptive changes no theory of adaptive evolution can be based purely on acquired changes." We point out that in natural selection there is no guarantee of adaptive changes either, and likewise, maladaptive changes are not always selected against. What can be said is that both natural selection and SOR possess mechanisms that amplify adaptive changes and diminish maladaptive ones (a point we elaborate on below).

Zachar et al. charge us with making a "baseline assumption to ignore maladaptive changes," but this is not the case. Figure 2 illustrates *adaptative response* of a RAF network to detrimental change. The network is not disrupted by the maladaptive disappearance of the stimulus that initiated its growth because it generates a foodset-derived element that triggers the reaction that was initially only triggered by that stimulus. This marks its transformation from a transient RAF (dependent on the continued presence of a particular outside stimulus) to a persistent RAF.

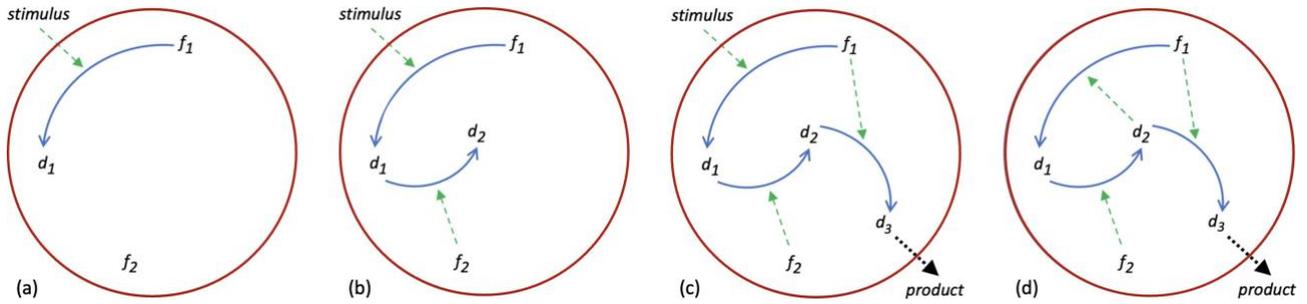

Figure 2. Growth and adaptation of a RAF network. $\{f_1, f_2\}$ are foodset elements; $\{d_1, d_2, d_3\}$ are foodset-derived elements; dashed green arrow denotes catalysis; blue arrow denotes reaction. (a) Stimulus catalyzes reaction resulting in $d_1$. (b) $f_2$ catalyzes reaction resulting in $d_2$. (c) $f_1$ catalyzes reaction that results in a product, $d_3$. (d) Because first reaction is catalyzed not just by stimulus but also by $d_2$, the RAF is no longer dependent on the stimulus. Adapted from [18].

We also do not accept Zachar et al.'s claim that evolution via SOR amounts to a "tautology". A tautology does not yield falsifiable predictions, yet as the commentators acknowledge, the SOR model predicts that complexity increases as the percolation parameter increases, and this was supported by their own simulations in their Supplementary Material (their Fig. 1). Similar 'tautology' claims have been made in the past concerning Darwinian natural selection (including by the philosopher of science Karl Popper, who later recanted this position), and that claim is now generally dismissed by biologists and philosophers [45, 46, 47, 56].

Contrary to Zachar et al.'s charge that our model is "irrelevant to any realistic context," we now list three examples of context-driven adaptive change in RAFs without selection: one from cognition / culture, one from artificial intelligence, and one from abiogenesis.

3.2.1. *Example from Culture / Cognition.* Our first example of context-driven adaptive change without selection comes from a paper that used RAFs to model how a psychotherapist elicits adaptive change in a client, illustrated in Figure 3 (For details see [31]). The therapist (Thera) offers perspectives that 'catalyze' new beliefs in the client (Clive), which help him function more adaptively. For example, she prompts him to find an alternative explanation for a bad grade that does not reinforce the negative self-belief 'I am stupid,' and she laughs at his joke, which catalyzes the positive self-belief 'I am funny.' Nowhere in this exchange is it necessary for Clive to express (i.e., actualize), all the different ways he could respond next, and for Thera to select amongst them. At each point in the exchange, there are multiple possible ways in which Clive's mental model of himself in relation to his world could have unfolded, and out of this potentiality, Thera draws upon her knowledge to elicit change in him that is more likely than chance to be adaptive. By altering how Clive's worldview is structured, Thera's influence is expected to infuse Clive's future interactions with others, and thereby impact his personal contribution to cultural evolution.

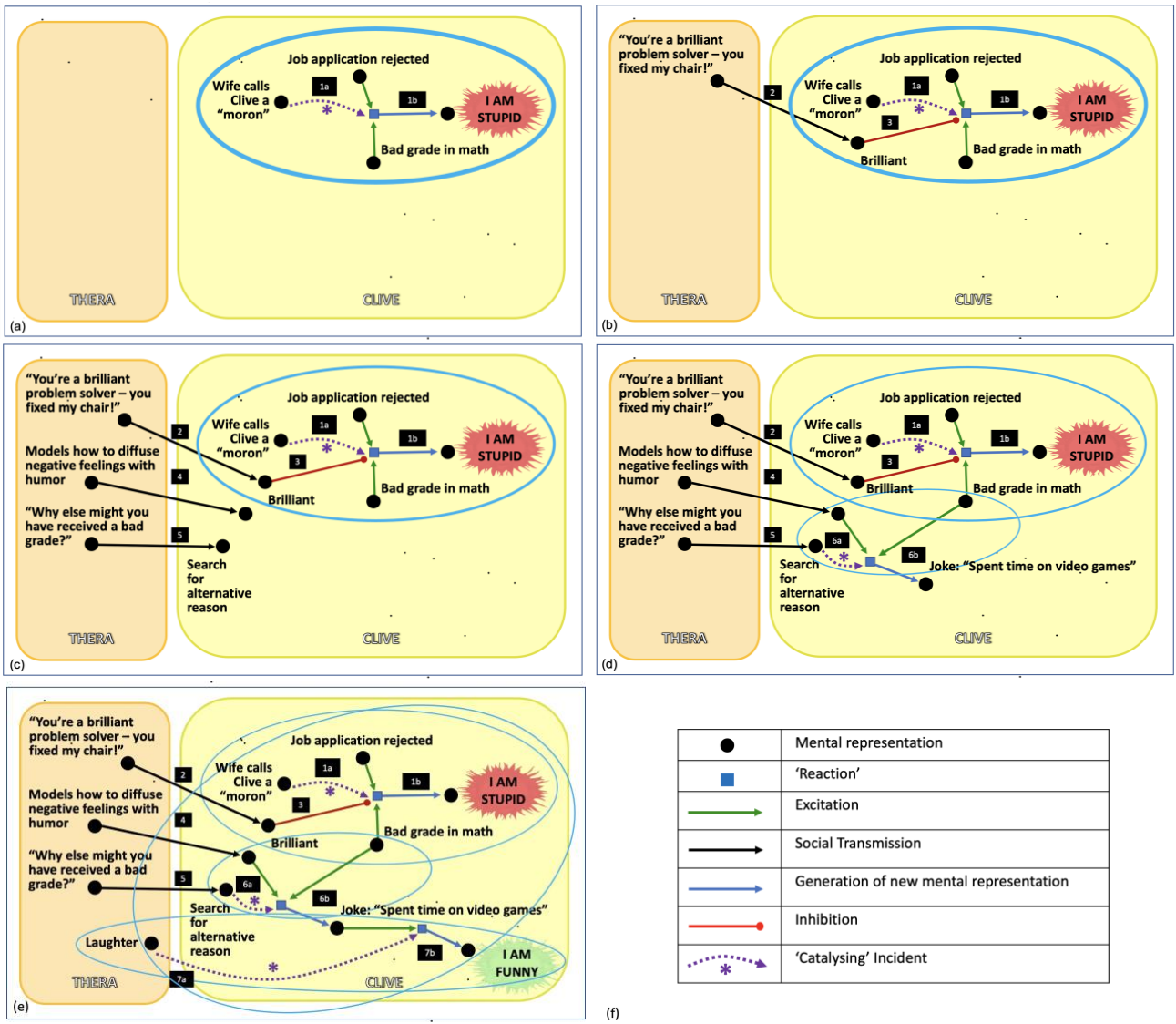

Figure 3. RAF model of how Clive's worldview is altered over the course of a psychotherapy session with Thera. (**a**) Following a 'catalyzing incident' in which his wife called him a "moron", Clive has been interpreting other events as confirmation of the distressing belief, 'I am stupid'. At the beginning of therapy these elements collectively constitute a stable RAF, as indicated by the thick blue line forming an oval around them. The thickness of this line indicates that the RAF has a large impact on Clive's thinking. (**b**) Thera praises Clive's brilliant problem solving ability, which generates a new foodset item, the notion that he is 'brilliant.' Since this is inconsistent with the belief 'I am stupid', it reduces the impact of that RAF, as indicated by thinning of the line that forms the blue oval. (**c**) Two more foodset items are transmitted from Thera to Clive. (**d**) Drawing upon Thera's modeling of how negative feelings can be diffused through humor, Clive makes a joke. The joke is catalyzed by Thera's prompt to explore alternate explanations of why he received a bad grade. When the joke diffuses negative feelings about the bad grade, it is less able to serve as a 'reactant' to support the 'I am stupid' belief, as illustrated by the further thinning of the oval representing that RAF. His joke constitutes a second RAF. (**e**) Thera's laughter at Clive's joke catalyzes a new belief, 'I am funny', which enhances his self-esteem, and forms a third RAF. These first three RAFs, which are irr-RAFs because they cannot be reduced further, interact and collectively form a maxRAF. (**f**) Key to symbols used in other panels. From [31]

3.2.2. *Example from Artificial Intelligence.* As a second example of context-driven adaptive change without selection, consider Large Language Models such as ChatGPT. Through iterative training on large datasets, by adjusting internal parameters to minimize errors they learn to recognize and predict patterns in how language gets used in different contexts. In this way, they exhibit adaptive change over time in their capacity for language comprehension and generation, not through natural selection (or anything algorithmically equivalent to it), but through statistical learning from this vast input data.

3.2.3. *Example from Abiogenesis.* RAF theory and methods have provided new insight into the formation of the kind of metabolism associated with the origin of life [60], and have shown how primitive self-replicating RNA systems generated in the lab can be described in terms of simpler autocatalytic networks [37]. The investigation of SOR in early life is more speculative than for cultural evolution due to a paucity of knowledge about early biochemistry. Nevertheless, the idea of primitive cell types evolving as a community (e.g. via horizontal transfer and/or by autocatalysis) prior to Darwinian evolution has been discussed by various authors (see e.g. [42, 52, 53, 55, 57, 58]).

3.3. **SOR is robust to degradation and imperfect replication and can accommodate detrimental stimuli and products, as well as maladaptive change.** In their analysis (Supplementary Material), Zachar et al. introduce processes and parameters ($\delta$, $\ell$) to represent 'degradation' and 'imperfect replication.' Their model deviates from the SOR model because it contains no RAF structure, and as such does not realistically portray how degradation and imperfect replication impact a self-organizing RAF. In RAF models, a RAF remains a RAF regardless of what additional elements and reactions are introduced to the system (though it may exhibit degradation in the dynamical sense that it is 'called upon' less often). In RAF structure, degradation need not be detrimental; for example, Figure 3 shows how 'degradation' of the initial RAF, due to inhibition from an external stimulus, paves the way for the emergence of a larger and more robust maxRAF. This example also shows that degradation in RAFs may involve inhibition, and we note that the RAF setting has been extended to include inhibition in abiogenesis [41] as well as cognitive [31] RAF models. Inhibition has yet to be be fully investigated in the SOR setting, but it has been shown that a reduction in the expected number of RAFs due to inhibition can be cancelled out by increasing the catalysis rate (see e.g., [41], Corollary 1).

Because Zachar et al's model contains no RAFs, it also does not capture the impact of imperfect replication in SOR, as is also evident from the example in Figure 3. Imperfect replication is a feature, not a drawback, of cultural evolution. For example, were the replication of cognitive structure from Thera to Clive—such as the knowledge that negative feelings can be diffused through humor—to be 'perfect,' it would have to include what originally catalyzed this knowledge in Thera (such as, perhaps, the experience of reading it in a textbook). However, this detail is irrelevant to Clive. Imperfect replication *ensures* that new information is assimilated in a way that is personally meaningful or relevant to existing cognitive structure in the recipient, which enables the recipient to contribute to culture by adapting this information to a new set of needs and tastes. (For further examples and discussion of imperfect replication of autocatalytic cognitive structure, see [8, 20, 21, 24, 26, 25, 28, 31, 32].)

Zachar et al. argue that SOR achieves a non-selectionist route to adaptation by artificially excluding detrimental stimuli and products.[3] However, SOR does not assume that stimuli and products are beneficial (or even that a stimulus or product *has* an a priori value prior to its interaction with the entity). Zachar et al. also falsely claim that SOR only allows for adaptive change. Indeed, Figure 3 depicts a non-beneficial stimulus (namely, the client's wife calling him a 'moron') catalyzing a maladaptive cognitive RAF (namely, the conclusion 'I am stupid'). The reason the impact of this maladaptive cognitive RAF eventually weakens is that the encouraging interactions with the therapist catalyze the formation of more robust cognitive RAF structure that is inconsistent with the maladaptive cognitive RAF. This example illustrates how, though detrimental elements are not excluded in SOR, they are less likely than beneficial elements to form stable RAF structure.

---

[3]For example, Zachar et al. write, "the authors' baseline assumption to ignore maladaptive changes so that they can leave out selection from the model."

Figure 1 in Zachar et al.'s commentary, intended to depict SOR, is not a model of SOR on several counts. First, it depicts no RAFs (no structures that meets the two criteria that define something as a RAF). Second, contrary to the phrase "randomly accumulating changes within SOR," changes in SOR do not accumulate randomly. At a given point in time, new change is both enabled and constrained by the structure of the existing network. Third, the figure depicts entity $k$ being selected out from a group of others, but in SOR, change occurs not by selection of the fittest but by transformation of all through the interleaving of self-organization and communal exchange, a process not depicted in this figure. Fourth, it is not the case that "only beneficial stimuli can reach the population," nor is it the case that SOR "only consider[s] adaptively beneficial products... and ignore[s] neutral or maladaptive ones." Fifth, it misapplies the notion of 'lotteries' which we discuss in the next section.

In essence, the RAF setting provides a means of modeling the process by which, through their interactions, a set of parts becomes a 'collective self', and there is little opportunity for elements that do not contribute to or reinforce that collective self to become part of it. Interestingly, the phenomenon of confirmation bias in psychology [44] shows that we *do* preferentially assimilate information that is consistent with our existing beliefs. This suggests there is psychological plausibility to the requirement in RAF models that new elements must 'latch into' existing network structure if they are to be accessible (i.e., able to be retrieved from memory).

3.4. **No Lotteries or Cherry-picking of Best-case Scenarios.** Zachar et al. claim that SOR involves three 'lotteries' and indicate this in their Fig. 1. They write, "A RAF of infinite adaptive potential is selected initially from many possible RAFs by unknown mechanisms (grey shapes, 1st lottery)." This description bears no relationship to what actually happens in SOR. First, although a RAF may have the potential to adapt in an infinite number of ways if exposed to an infinite number of contexts, we have never claimed that SOR involves a RAF of "infinite adaptive potential." Second, SOR does not involve selection of one RAF from many possible RAFs; rather, it involves emergence of a RAF by way of catalyzed interactions amongst components. The mechanism by which this happens are not "unknown"; they are described in Section 2 of [27], and at length in dozens of other papers, many of which were cited in [27].

Zachar et al. write, "the authors assume only beneficial environmental stimuli (lottery #2) and beneficial products (lottery #3)." Section 3.3 dealt at length with why SOR does not assume that either stimuli or products are beneficial. A further misunderstanding is revealed in the wording of this sentence about what stimuli and products actually are. First, it does not make sense to refer to a stimulus or product except with respect to a particular individual; one individual's product is another individual's stimulus. Second, it makes no sense to refer to a stimulus as inherently beneficial or detrimental; a stimulus that is beneficial to one individual may be detrimental to another. Indeed, the same stimulus (or product) may be simultaneously beneficial to an individual in one respect, and detrimental to that same individual in another respect. To go back to the example in Figure 3, the stimulus that catalyzed a maladaptive cognitive RAF in Clive (his wife calling him a 'moron') was a product with respect to his wife, and could have been perceived by her as beneficial in that it enabled emotional release.

Zachar et al. write, "While the best-case scenario the authors incorrectly generalize from may be mathematically valid, stripping away their unrealistic assumptions reveals that SOR does not represent real entities (e.g., protocells) but rather models the triviality that fast horizontal diffusion of effects can effectively equalize a population." The last part of this sentence seems to miss a key result in our model—increasing the rate of transfer leads to increased complexity—a result they reproduced in their model when applied to our setting (Fig. 1 of their Supplementary Material). There, they write "We reproduced the SOR model and the central claim of [27] that it can implement evolution without variation. We were able to reproduce the major claims of the original paper". Their further analysis introduced a degradation process, but did not incorporate actual RAF structure, and therefore did not incorporate that a self-sustaining RAF grows in such a way that its robustness against changes that would harm or degrade it increases with time.

Zachar et al. accuse us of starting from cherry-picking a "right RAF," but our examples do not start with a "right RAF", indeed, most of our examples start off with no RAF at all. The point of our examples is to illustrate how RAFs self-organize from discrete *parts* (be they catalytic molecules, or interacting mental representations) that do not have the capacity to self-replicate and evolve, to *wholes* that do.

### 3.5. **SOR is Not Merely "Akin to Learning".**

Zachar et al.'s claim that "SOR is a process akin to learning" is misleading, as models of learning do not set out to explain how, through their interactions, a set of parts become a self-replicating, evolving whole. [27] is explicit about the relationship between SOR and learning; indeed Section 4.1 is entirely devoted to this topic. In RAF models, which are the core of SOR, the learning or assimilation of new information is mathematically described by the incorporation of new *foodset* items. Conversely, creative reprocessing and accommodation of existing knowledge in response to new information is mathematically described by the generation of *foodset-derived* items. In short, SOR includes not just the learning of new knowledge, but also the creative reprocessing and accommodation of existing knowledge. It includes how the fruits of learning and creative thought percolate not just between individuals of a population, but throughout the cognitive network of each individual in the population.

### 3.6. **SOR is not a pure percolation model.**

It is particularly odd that, though Zachar et al. claim that "SOR is a process akin to learning...," they repeatedly refer to SOR as a percolation model, despite that percolation models do not incorporate learning (nor creative processes, nor the self-organization of parts into a collective whole). Zachar et al. write, "A model of infection is not a model of evolution," and we wholeheartedly agree; this is in essence a point one of us has been making in the cultural evolution community for decades [9, 10, 11, 12, 13, 14, 15, 16, 17, 54].[4] It seems fair to say that RAF-based models of cultural evolution (cited in Section 2.3) take internal cognitive processes into account to a much greater degree than most (e.g., [1, 3, 4, 5, 51]) though not all (c.f. [2, 6]) other contemporary approaches.

In Zachar et al.'s vision of SOR, "External stimuli may trigger the generation of new beneficial products in a population member that ultimately percolate to every entity causing thus community-based adaptation without reproduction, birth, death, or competition, i.e., without Darwinian dynamics." However, this kind of approach to culture is exactly what SOR was developed to replace. SOR emphasizes what goes on within individuals as opposed to what goes on between them. Zachar et al. equate 'products' in SOR with 'new infections' in a percolation model, but we are at a loss to understand this analogy. A product is created *by* the individual, and exists *outside* that individual; however, neither is the case for a new infection. Figure 2 in Zachar et al.'s commentary incorrectly designates SOR as a pure percolation-based model.

3.6.1. *Comparison Between SOR and SIR.* Zachar et al. compare SOR to SIR, a specific percolation model that (in their words) "assumes three types of individuals: susceptible (who may acquire the infection), infected (who got the infection) and recovered (who are cleared of the infection or are dead) and omits birth and death dynamics." They claim, "SOR is practically the simplest case of SIR, assuming no recovery." However, other than their acronyms, we see virtually no similarity between SOR and SIR.

---

[4]The cultural evolution community has long assumed that cultural variation arises through simplistic phenomena such as 'prestige bias' and 'replication error,' paying only lip service to the complex assimilation/accommodation process by which information is encoded, and the iterative creative process by which new information is gener- ated. A related problem in the cultural evolution literature is the tendency to equate individual learning (i.e., the acquisition of of *pre-existing* information, which in the RAF approach is modeled as foodset elements) with creativity (i.e., the generation of *new* information, which in the RAF approach is modeled as foodset-derived elements).

First, SOR and SIR set out to explain different phenomena. SOR is not a model of infection; it is a model of how collective wholes formed by self-organized interactions amongst their parts can exhibit cumulative, adaptive change.

It is true that sometimes models developed to explain different phenomena may be found to converge with respect to their underlying mechanisms. However, this is not the case with respect to SOR and SIR. There are no collective wholes in SIR, and no cumulative, adaptive change in SIR. (Zachar et al. write, "successive infections in SIR cannot be considered a model of evolution," and with this we agree.)

Oddly, even the charge that SOR "assum[es] no recovery," is incorrect. We again refer the reader to Figure 3, which illustrates a RAF model of recovery from a detrimental stimulus. By the last panel, the client's maladaptive belief is still present in his memory, but it is no longer a debilitating threat to his psychological wellbeing (he has 'recovered' from it), and he is able to function in an adaptive way. The figure does not show how this cognitive change affected the client's subsequent interactions with others, but were this included, it could be considered a SOR model.

In short, SOR and SIR are completely different both with respect to what they set out to explain, and with respect to their mechanisms.

## 4. Why we Model a Population Instead of a Single RAF

Zachar et al. write, "we do not understand why it was important at all for Gabora and Steel to model a population of RAFs instead of a single RAF."

There are several reasons we model a population instead of a single RAF. First, although the union of RAFs is a RAF, this amalgamation introduces complications that were irrelevant to the basic arguments of the paper. For example, in a RAF model of cultural evolution, taking a union requires making decisions about whether a MR in the mind of one individual is to be regarded as the same MR in another person's mind.

In addition, we were comparing SOR with natural selection, which occurs at the level of populations, as illustrated in Fig. 2 of [27]. The best way to compare two things is to hold all else constant and see what differs. Readers would not have been able to understand how SOR compares with natural selection if we had compared what happens at the population level in natural selection with what happens at the individual level in SOR.

Yet another reason it was necessary to look at the population level is that there are nonlin- ear interactions between the internal (self-organization) component of SOR and the external (transmission) component. This was evident even in the simple agent-based models that the first author developed in the 1990s [7], which served as a precursor to subsequent models of cultural evolution, as well as our current RAF-based models. Fig. 4 shows that when the agents *only* imitated—i.e., 100% external transmission *between* agents—there was no cultural evolu- tion at all. When they *only* invented—i.e., 100% creation *within* agents—the mean fitness of cultural outputs across the artificial society was sub-optimal. It only approached optimality when there was a balance between inventing and imitating. Following this initial experiment, there have been other more in-depth studies of this phenomenon [22, 29, 30].

In short, whether it be a model of natural selection, SOR, or even this simple 'SOR-precursor' model of cultural evolution, it is not possible to fully understand the scope of individual-level changes, and their impact on cumulative, adaptive change (i.e., what we call evolution) without looking at the level of the population.

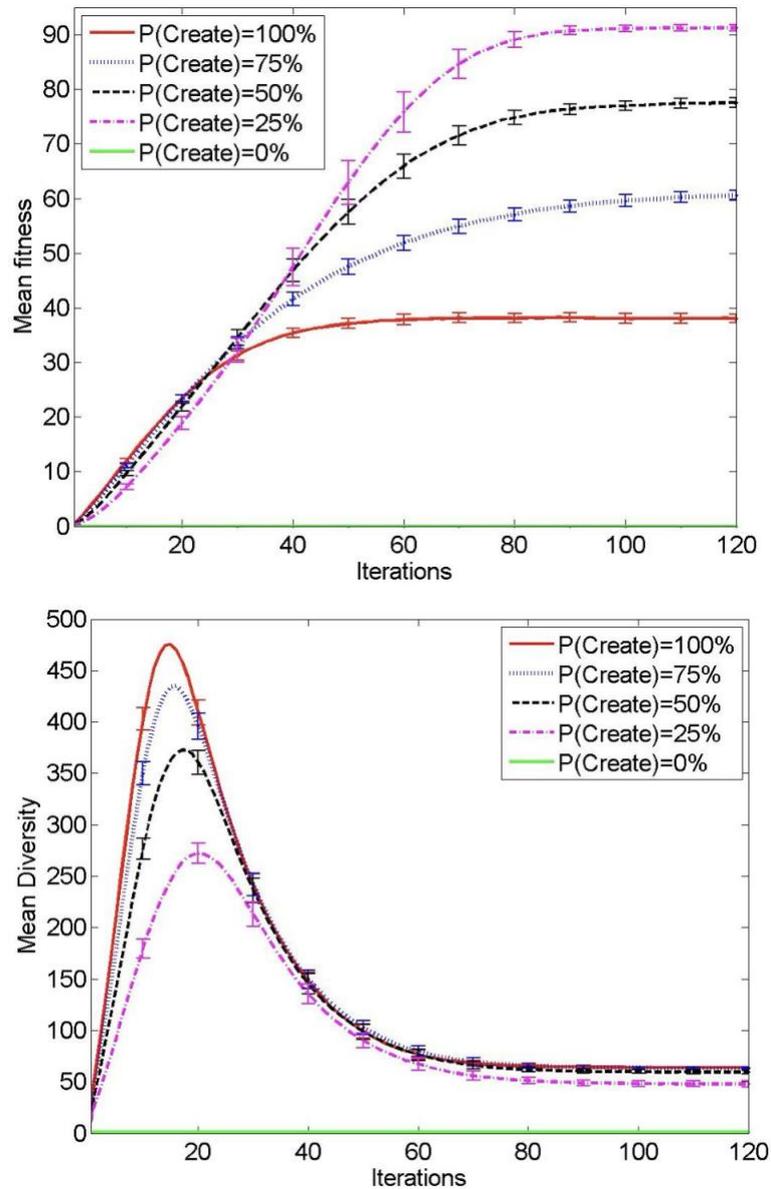

Figure 4. Fitness (top) and diversity (bottom) of cultural outputs with different ratios of inventing to imitating. From [15].

5. **On the Inevitability of Darwinian Dynamics**

Zachar et al. write, "if (membrane-bounded) RAFs grow autocatalytically, it should inevitably lead to cellular and populational dynamics inevitably leading to competition, birth and death and differential survival or replication of individuals, i.e., to conventional Darwinian dynamics." We have no basis upon which to speculate on the inevitability of RAF dynamics giving way to Darwinian dynamics, and suggest that the onus falls on the commentators to provide evidence for their claim that this is inevitable.[5] In any case, since no claims were made regarding the permanence of a SOR process in [27], this issue is not directly relevant to the arguments in that paper.

## 6. **Randomness**

Zachar et al. write:

> [A] RAF "tumbling down" a sequence of truly random environmental stimuli, not only beneficial ones, would be as adaptive as a rock tumbling down a stream. In reality, random effects are more likely to be neutral or maladaptive and there is absolutely no guarantee that newly generated products of a RAF are beneficial.

However, neither is randomness a necessary precondition for SOR to be applicable, nor is it the case that the environmental stimuli involved in the emergence of RAF structure are random. In the case of the application of RAFs to the emergence of integrated cognitive structure and cumulative cultural change, stimuli are most certainly not random. Parents go out of their way to expose their offsprings' developing minds to exactly those stimuli that catalyze productive conceptual change, as discussed and modeled in [21]. Internal catalysts of conceptual change are even less random than external catalysts of conceptual change, as discussed and modeled in [28]. Minds generate novelty (in RAF terms, foodset-derived elements) using not just strategy, but intuitions born of pattern recognition and pattern completion on vast quantities of data.

_______________

[5]In discussion with Kalin Vetsigian, first author of a paper that presented a computational model of the transition to what they refer to as the 'Darwinian threshold' [55], he speculated that on Earth, the transition from nonDarwinian to Darwinian evolution took on the order of 100 million years.

## 7. **Survival of *Fittest* versus Context-driven Adaptive Change of *All***

Zachar et al. write that competition, birth and death and differential survival or replication of individuals, i.e., conventional Darwinian dynamics "are entirely ignored in SOR to maintain the claim that it is a populational and evolutionary process without population dynamics." It is not that these processes are *ignored* in SOR, it is that population dynamics occurs through a different mechanism. Instead of competitive exclusion, i.e., survival and replication of the fittest at the expense of the less fit, it works through transformation of all, as illustrated in Fig. 2 of [27].

Change in natural selection works through *selection on actualized entities;* multiple entities exist in well-defined states in the physical world, and some survive and reproduce while others do not. Change in SOR can work through a different mechanism: *context-induced actualization of potential*, [19, 23]. An entity in a state $p(t_i)$ at an instant of time ($t_i$) has the potential to change to multiple other states $p_1(t_i + 1), p_2(t_i + 1), ..., p_n(t_i + 1), ....$ None of these states are (yet) actual; they are all merely potential. (We invite the commentators to think of it this way: you may insist that there *is* variation, but it is present not in the sense that it is real, or exists, but in the sense that it is possible; it has the potential to exist.) Which of its potential future states $p_1(t_i + 1), p_2(t_i + 1), ..., p_n(t_i + 1), ...$ the entity actually does change to depends on the contexts it interacts with. Each successive context it encounters contains information that influences the entity's change of state at that instant.[6]

## 8. **Conclusions**

We appreciate that the authors of the commentary have taken the time to lay out their thoughts on our paper [27], and we are grateful to the editors for the opportunity to write this reply to the commentary. We hope it will help the commentators to see why, once common terminology is established and misconceptions are cleared up, the points made in the commentary do no damage to the arguments laid out in [27]. RAF theory has proven useful for modeling biochemical and metabolic processes relevant to both the origin of life, and the origin of cognitive structures capable of cumulative, adaptive cultural change.

---

[6]There may be multiple contexts that the entity could interact with at $t_i + 1$, which we can denote $e_1(t_i + 1), e_2(t_i + 1), , e_n(t_i + 1),$  To determine the state of the entity at $t_i + 1$ we must know the state of the context it interacts with at $t_i + 1$, and we may not be able to describe the state of that context, or how it interacts with that entity with sufficient precision. If the change of state is nondeterministic, we cannot predict with complete certainty the state of the entity at $t_i + 1$.

If the commentators are resolutely intent on insisting that variation and selection are present in SOR, there is a sense in which this could be said to be the case. The multiple potential future states of an entity might be considered 'variants,' and one might say the context 'selects' one of these potential states to become actual. This, however, is not what is conventionally meant by the terms 'variation' and 'selection,' and context-driven change of state (which encompasses both learning and creativity), requires a different mathematical framework from that of natural selection [19]. In SOR, it isn't that adaptive as well as non-adaptive entities are generated, and exist in the real world, and the non-adaptive ones are subsequently weeded out. Instead, out of the many potential future states of an entity, the context elicits a state that is more likely than chance to be adaptive with respect to that context.

## 9. Acknowledgements

This research was funded by Grant 62R06523 from the Natural Sciences and Engineering Research Council of Canada.

## References

[1] E. Boon, P. van den Berg, L. Molleman, and F. J. Weissing, *Foundations of cultural evolution*, Philosophical Transactions of the Royal Society B, 376 (2021), p. 20200041.

[2] A. Buskell, M. Enquist, and F. Jansson, *A systems approach to cultural evolution*, Palgrave Communications, 5 (2019), pp. 1–15.

[3] N. Creanza, O. Kolodny, and M. W. Feldman, *Cultural evolutionary theory: How culture evolves and why it matters*, Proceedings of the National Academy of Sciences, 114 (2017), pp. 7782–7789, `https://doi.org/https://doi.org/10.1016/j.shpsc.2016.12.005`.

[4] D. Deffner, A. Kandler, and L. Fogarty, *Effective population size for culturally evolving traits*, PLoS Computational Biology, 46 (2022), pp. 1–22, `https://doi.org/https://doi.org/10.1371/journal.pcbi.1009430`.

[5] M. Derex and R. Boyd, *Partial connectivity increases cultural accumulation within groups*, Proceedings

of the National Academy of Sciences of the United States of America, 113 (2016), pp. 2982–2987, `https://doi.org/https://doi.org/10.1073/pnas.1518798113`.

[6] J. B. Falandays and P. E. Smaldino, *The emergence of cultural attractors: how dynamic populations of learners achieve collective cognitive alignment*, Cognitive Science, 46 (2022), p. e13183, `https://doi.org/https://doi.org/10.1016/j.shpsc.2016.12.005`.

[7] L. Gabora, *Meme and variations: A computer model of cultural evolution*, in 1993 Lectures in Complex Systems, A. Goel, C. Seifert, and C. Freska, eds., Addison-Wesley, Boston MA, 1995, pp. 471–486.

[8] L. Gabora, *Autocatalytic closure in a cognitive system: A tentative scenario for the origin of culture*, Psycoloquy, 9 (1998), pp. [adap–org/9901002].

[9] L. Gabora, *Cognitive mechanisms underlying the origin and evolution of culture*, doctoral dissertation, Free University of Brussels, Belgium, 2001.

[10] L. Gabora, *Ideas are not replicators but minds are*, Biology and Philosophy, 19 (2004), pp. 127–143.

[11] L. Gabora, *The fate of evolutionary archaeology: Survival or extinction?*, World Archaeology, 38 (2006), pp. 690–696.

[12] L. Gabora, *The cultural evolution of socially situated cognition*, Cognitive Systems Research, 9 (2008), pp. 104–113.

[13] L. Gabora, *Five clarifications about cultural evolution*, Journal of Cognition and Culture, 11 (2011), pp. 61–83.

[14] L. Gabora, *An evolutionary framework for culture: Selectionism versus communal exchange*, Physics of Life Reviews, 10 (2013), pp. 117–145, `https://doi.org/10.1016/j.plrev.2013.03.006`.

[15] L. Gabora, *Handbook of culture and creativity: Basic processes and applied innovations*, in . The creative process of cultural evolution, A. Leung, ed., Oxford University Press, New York, 2018, pp. 33–60.

[16] L. Gabora, *Creativity: Linchpin in the quest for a viable theory of cultural evolution*, Current Opinion in Behavioral Sciences, 27 (2019), pp. 77–83, `https://doi.org/10.1016/j.cobeha.2018.09.013`.

[17] L. Gabora, *From deep learning to deep reflection: Toward an appreciation of the integrated nature of cognition and a viable theoretical framework for cultural evolution*, in Proceedings of Meeting of the Cognitive Science Society, L. Nadel and D. Stein, eds., Cognitive Science Society, Austin, TX, 2019, pp. 1801–1807.

[18] L. Gabora, *How does culture evolve?*, in 10th Volume of the Annual Series on Advances in Culture and Psychology, M. Gelfand, C. Chiu, and Y. Hong, eds., Oxford University Press, New York, 2024.

[19] L. Gabora and D. Aerts, *Evolution as context-driven actualisation of potential: toward an interdisciplinary theory of change of state*, Interdisciplinary Science Reviews, 30 (2005), pp. 69–88.

[20] L. Gabora and J. Bach, *A path to generative artificial selves*, in Proceedings of the 22nd Portuguese Conference on Artificial Intelligence, N. Moniz, Z. Vale, J. Cascalho, C. Silva, and R. Sebastiao, eds., Springer- Verlag, Berlin, 2023, pp. 15–29, `https://doi.org/https://doi.org/10.1007/978-3-031-49011-8_2`.

[21] L. Gabora, N. Beckage, and M. Steel, *An autocatalytic network model of conceptual change*, Topics in Cognitive Science (Special issue edited by Y. Kenett and T. Hills on 'Networks of the Mind: How can Network Science Elucidate Our Understanding of Cognition?'), 14 (2022), pp. 163–188, `https://doi.org/`

https://doi.org/10.1111/tops.12583.

[22] L. Gabora and H. Firouzi, *Society functions best with an intermediate level of creativity*, in Proceedings of the 34th Annual Meeting of the Cognitive Science Society, N. Miyake, D. Peebles, and R. P. Cooper, eds., Cognitive Science Society, Austin TX, 2012, pp. 1578–1583.

[23] L. Gabora and S. Kauffman, *Toward an evolutionary-predictive foundation for creativity*, Psychonomic Bulletin & Review, 23 (2016), pp. 632–639, https://doi.org/10.3758/BF03196761.

[24] L. Gabora and M. Steel, *Autocatalytic networks in cognition and the origin of culture*, Journal of Theoretical Biology, 431 (2017), pp. 87–95, https://doi.org/10.1016/j.jtbi.2017.07.022.

[25] L. Gabora and M. Steel, *A model of the transition to behavioral and cognitive modernity using reflexively autocatalytic networks*, Proceedings of the Royal Society Interface, 17 (2020), p. 20200545, https://doi.org/http://doi.org/10.1098/rsif.2020.0545.

[26] L. Gabora and M. Steel, *Modeling a cognitive transition at the origin of cultural evolution using autocatalytic networks*, Cognitive Science, 44 (2020), p. e12878.

[27] L. Gabora and M. Steel, *An evolutionary process without variation and selection*, Journal of The Royal Society Interface doi: 10.1098/rsif.2021.0334, 18 (2021), p. 20210334.

[28] L. Gabora and M. Steel, *From uncertainty to insight: An autocatalytic framework*, in Uncertainty: A Catalyst for Creativity, Learning and Development, R. Beghetto and G. Jaeger, eds., Springer, Berlin, 2022, pp. 125–158.

[29] L. Gabora and S. Tseng, *The social impact of self-regulation on the evolution of simple and complex creative ideas*, in Proceedings of the 5th International Conference on Computational Creativity, S. Colton, M. Cook, N. Lavrac, and D. Ventura, eds., Association for the Advancement of Artificial Intelligence (AAAI) Press, Palo Alto, 2014, pp. 550–558.

[30] L. Gabora and S. Tseng, *The social benefits of balancing creativity and imitation: Evidence from an agent-based model*, Psychology of Aesthetics, Creativity, and the Arts, 11 (2017), pp. 457–473.

[31] K. Ganesh and L. Gabora, *A dynamic autocatalytic network model of therapeutic change*, Entropy (Special issue edited by W. Tschacher & F. Orsucci on 'Complexity Science and Human Change.'), 44 (2022), p. 547, https://doi.org/https://doi.org/10.3390/e24040547.

[32] K. Ganesh and L. Gabora, *Modeling discontinuous cultural evolution: The impact of cross-domain transfer*, Frontiers in Psychology - Theoretical and Philosophical Psychology, 13 (2022), p. 786072, https://doi.org/10.3389/fpsyg.2022.786072.

[33] W. Hordijk, J. Hein, and M. Steel, *Autocatalytic sets and the origin of life*, Entropy, 12 (2010), pp. 1733–1742, https://doi.org/10.3390/e12071733.

[34] W. Hordijk, S. A. Kauffman, and M. Steel, *Required levels of catalysis for emergence of autocatalytic sets in models of chemical reaction systems*, International Journal of Molecular Science, 12 (2011), pp. 3085–3101, https://doi.org/10.3390/ijms12053085.

[35] W. Hordijk and M. Steel, *Detecting autocatalytic, self-sustaining sets in chemical reaction systems*,

Journal of Theoretical Biology, 227 (2004), pp. 451–461, `https://doi.org/10.1016/j.jtbi.2003.11.020`.

[36] W. Hordijk and M. Steel, *Predicting template-based catalysis rates in a simple catalytic reaction model*, Journal of Theoretical Biology, 295 (2012), pp. 132–138.

[37] W. Hordijk and M. Steel, *A formal model of autocatalytic sets emerging in an RNA replicator system*, Journal of the Systems Chemistry, 4:3 (2013), `https://doi.org/10.1186/1759-2208-4-3`.

[38] W. Hordijk and M. Steel, *Autocatalytic sets and boundaries*, J. Syst. Chem., 6:1 (2015).

[39] W. Hordijk and M. Steel, *Chasing the tail: The emergence of autocatalytic networks*, Biosystems, 152 (2016), pp. 1–10, `https://doi.org/10.1016/j.biosystems.2016.12.002`.

[40] W. Hordijk, M. Steel, and P. Dittrich, *Autocatalytic sets and chemical organizations: Modeling self-sustaining reaction networks at the origin of life*, New Journal of Physics, 20 (2018), p. 015011.

[41] S. Kauffman and M. Steel, *The expected number of viable autocatalytic sets in chemical reaction systems*, Artificial Life, 27 (2021), pp. 1–14.

[42] S. A. Kauffman, *The origins of order*, Oxford University Press, 1993.

[43] E. Mossel and M. Steel, *Random biochemical networks and the probability of self-sustaining autocatalysis*, Journal of Theoretical Biology, 233 (2005), pp. 327–336, `https://doi.org/10.1016/j.jtbi.2004.10.011`.

[44] R. S. Nickerson, *Confirmation bias: A ubiquitous phenomenon in many guises*, Review of General Psychology, 2 (1998), pp. 175–220.

[45] E. S. Reed, *The lawfulness of natural selection*, Am. Nat., 118 (1981), pp. 61–71.

[46] E. Sober, *Philosophy of Biology*, Routledge, 2018.

[47] E. Sober, *The Philosophy of Evolutionary Theory: Concepts, Inferences and Probabilities*, Cambridge University Press, 2024.

[48] M. Steel, W. Hordijk, and J. C. Xavier, *Autocatalytic networks in biology: Structural theory and algorithms*, Journal of the Royal Society Interface, 16 (2019), p. rsif.2018.0808, `https://doi.org/10.1098/rsif.2018.0808`.

[49] M. Steel, J. C. Xavier, and D. H. Huson, *Autocatalytic networks in biology: Structural theory and algorithms*, Journal of the Royal Society Interface, 17 (2020), p. 20200488.

[50] L. Steels and E. Szathmáry, *The evolutionary dynamics of language*, Biosystems, 164 (2018), pp. 128–137, `https://doi.org/https://doi.org/10.1016/j.biosystems.2017.11.003`, `https://www.sciencedirect.com/science/article/pii/S0303264717302885`. Code Biology.

[51] K. Sterelny, *Cultural evolution in california and paris*, Studies in History and Philosophy of Science Part C: Studies in History and Philosophy of Biological and Biomedical Sciences, 62 (2017), pp. 42–50, `https://doi.org/https://doi.org/10.1016/j.shpsc.2016.12.005`.

[52] V. Vasas, C. Fernando, M. Santos, S. Kauffman, and E. Szathmáry, *Evolution before genes*, Biology Direct, 7 (2012).

[53] V. Vasas, C. Fernando, A. Sziágyi, I. Zachár, and E. Santos. M. Szathmáry, *Primordial evolvability: impasses and challenges*, Theoretical Biology, 381 (2015).

[54] T. Veloz, I. Temkin, and L. Gabora, *A conceptual network-based approach to inferring the cultural evolutionary history of the baltic psaltery*, in Proceedings of the 34th Annual Meeting of the Cognitive Science Society, N. Miyake, D. Peebles, and R. Cooper, eds., Cognitive Science Society, Austin TX, 2012, pp. 2487–2492.

[55] K. Vetsigian, C. Woese, and N. Goldenfeld, *Collective evolution and the genetic code*, Proceedings of the National Academy of Sciences USA, 103 (2006), pp. 10696–10701.

[56] M. E. Weale, *Darwinism not tautological*, Nature, 351 (1991), p. 600.

[57] M. P. Williamson, *Autocatalytic selection as a driver for the origin of life*, Life, 14 (2024), pp. 1–17, https://doi.org/10.3390/life14050590.

[58] C. R. Woese, *On the evolution of cells*, Proceedings of the National Academy Science, 99 (2002), pp. 8742–8747.

[59] J. C. Xavier, W. Hordijk, S. Kauffman, M. Steel, and W. F. Martin, *Autocatalytic chemical networks at the origin of metabolism*, Proceedings of the Royal Society of London. Series B: Biological Sciences, 287 (2020), p. 20192377.

[60] J. C. Xavier and S. Kauffman, *Small-molecule autocatalytic networks are universal metabolic fossils*, Phil. Trans. Roy. Soc. A, 380 (2022).

[61] I. Zachar, M. Máté, and S. Számadó, *Tautology explains evolution without variation and selection. a comment on: 'an evolutionary process without variation and selection'*, Journal of the Royal Society Interface, 21 (2024), p. 20210334, https://doi.org/https://doi.org/10.1098/rsif.2021.0334.